\documentclass[twocolumn,aps, prd]{revtex4}%
\usepackage{graphicx}
\usepackage{dcolumn}
\usepackage{bm}
\usepackage{amsmath}%
\usepackage{amsfonts}%
\usepackage{amssymb}

\begin{document}

\title{Possible evidence of a ground level enhancement of muons in association with a SWIFT Trigger}

\author{C. R. A. Augusto, C. E. Navia\footnote{e. mail: navia@if.uff.br}, M. B. Robba and K. H. Tsui}
\address{Instituto de F\'{\i}sica Universidade Federal Fluminense, 24210-346,
Niter\'{o}i, RJ, Brazil} 

\date{\today}
\begin{abstract}
Starting from April 2007, a search for solar daily variation of the muon intensity ($E_\mu >0.2$ GeV) at sea level and using two directional muon telescopes is in progress. In this survey, several ground level enhancements (GLEs)
on the muon counting rate background have been found. Here, we highlight one of them, observed in the vertical telescope on 07 August 2007 for the following reasons: The GLE consists of a single narrow peak, with a statistical significance of 4.4$\sigma$.
The GLE is in temporal coincidence with a SWIFT trigger $N^0287222$, at 21:16:05 UT according to the Burst Alert Telescope (BAT) on board of the SWIFT spacecraft. However, the Swift StarTracker had lost stellar lock minutes before that and the resulting improper s/c attitude information caused BAT
to "trigger" on a known source. Even so, the SWIFT trigger coordinates are inside the effective field of view of the vertical Tupi muon telescope. The temporal and directional coincidences between this GLE and the SWIFT satellite unknown event strongly suggest that they may be physically associated. Details and implications of this possible association are reported in this work. 
\end{abstract}


\maketitle

\section{Introduction}

Gamma Ray Bursts (GRBs) are among the most powerful astrophysical phenomena in the Universe. They were discovered by accident during the cold war years by American military satellites designed to detect gamma rays produced by Soviet atomic bomb test on Earth. The GRBs appear first as a brilliant flash of gamma rays, that rises and falls in a matter of minutes. These bursts are often followed by afterglows at X-ray, optical and radio wavelengths. The Burst and Transient Source Experiment (BATSE), on board of the Compton spacecraft \cite{meegan96}, had detected thousands of GRBs in a period of 9 years, about a per day. In addition the BATSE results had shown that GRBs occurred at random all over the sky \cite{paciesas99}. The first measurements of the redshifts in GRB afterglows by Beppo SAX \cite{costa97}, together with the highly isotropic distribution of their arrival directions, have established their cosmological origin, at least for the GRBs of long duration. 

A statistical survey has shown that it is not always possible to spectroscopically determine the red shift even in some long well located GRBs accompanied by X-ray, optical and radio afterglows. A plausible explanation for these results is the assumption of a local origin. That is to say, they are close to our Galaxy or inside it. 
Around 70 percent of the observed GRBs are the long-soft type with approximately 30 s duration. Some of these bursts have their measured red shifts clustering in $z=1$, consequently they have a cosmological origin and are connected with supernovae events. However, there is also another category of GRBs, the short-hard type with approximately 0.2 s duration, where no red shift and no afterglow have been detected from these bursts. Probably these short bursts result from a different engine than the long bursts. They could be connected with compact binary mergers, like two neutron stars or with a black hole component, as well as binary pulsar.

The enhancements of particles at ground level from GRBs remain an open issue, at least, they have not been observed with a high confidence level. The Milagrito experiment, which is a predecessor of the water Cerenkov MILAGRO experiment, has reported evidences of the TeV counterpart of a BATSE GRB (trigger 970417) \cite{atkin00}. The GRAND project (muon detector at ground level) has reported also some evidences of GRB detection in coincidence with one BATSE GRB (Trigger 971110), although with a low significance level \cite{pourier03}. The MILAGRO detector has a wide field of view ($\sim 2 sr$) and high duty cycle ($>90\%$). MILAGRO has reported the $>5\sigma$ discovery of a VHE gamma excess from the inner galaxy. In addition, in the last two International Cosmic Ray Conference (Pune 2005 and Merida 2007), search for short duration VHE emission from GRBs with MILAGRO detector has been reported \cite{vasileiou07}, and
no significant evidence was observed for VHE emission from GRBs with duration ranging from 250 $\mu s$ to $40$ s in 2.3 years of observation \cite{abdo07}. In addition, in the same conferences other ambitious projects for GeV GRBs detection at ground have been reported such as the ARGO air shoer project at Tibet and the STACEE and MAGIC, two atmospheric Cherenkov detectors. In fact, the MAGIC Group has reported the observation of the GRB050713a, triggered by a SWIFT alert. However, they conclude that using standard MAGIC  analysis, no evidence for a gamma signal was found
\cite{albert06}.  

On the other hand, several models for the origin of GRBs predicts a TeV component \cite{totani96,dermer00,pilla98}. A plausible explanation for the
extremely discrepancy between the observations and predictions in the GeV-TeV region is to invoke the attenuation of TeV photons by interaction with the intergalactic  infrared and light radiations. This process would responsible for a cutoff for TeV GRBs, for example with a detector with a threshold energy of 1.0 TeV is not possible to observed sources with $z\geq 0.1$. An energy threshold of 300 GeV allows to reach $z\approx 0.2$, and a threshold of 100 GeV allows to reach $z\approx 1.0$.  This means that ground based detectors, such as Tupi, observing muons have an open universe up to $z\approx 0.2-0.5$. 

The EGRET detection of GRBs up to 18 GeV \cite{hurley94} and the spectra of several GRBs above 100 MeV gave no indications of any spectral attenuation that might preclude detection of bursts at higher energies. These features have encouraged the use of ground based methods to detect GRBs in coincidence with satellites.

\section{The Tupi experiment}

\begin{figure}[th]
\vspace*{-2.0cm}
\includegraphics[clip,width=0.5
\textwidth,height=0.5\textheight,angle=0.] {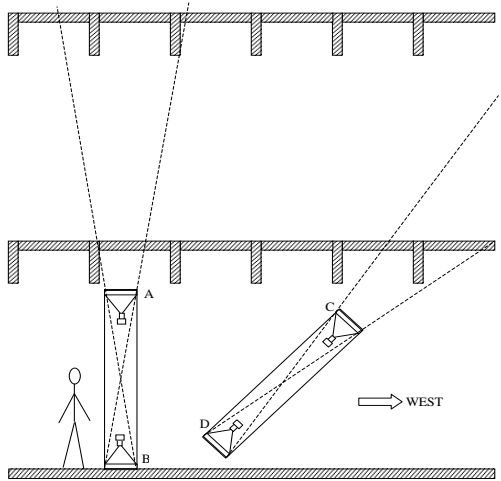}
\vspace*{-4.0cm}
\caption{Experimental setup of the Tupi experiment Phase II, showing the two telescopes.}
\end{figure}

The Tupi experiment, located at sea level with coordinates $S22^054'33''$ latitude and $W43^008'39''$ longitude, was a muon tracking telescope on the basis of scintillator detectors. Results on correlations between solar transient events and Tupi muon flux on ground, as well as details on the experimental setup can be found in earlier publications \cite{augusto03,navia05,augusto05}.
Starting from April 2007, we have initiated the Phase II of the Tupi experiment with two identical muon telescopes on the basis of plastic scintillators. One of them with vertical orientation and the other with an orientation of 45 degrees of the vertical (zenith) and pointing to the west. Both with an effective aperture of $65.6\;cm^2\;sr$. The telescopes are inside a building under two flagstones of concrete, allowing registration of muons with $E_{\mu}\geq 0.2\;GeV$ required to penetrate the two flagstones, as shown in Fig.1.
The rigidity of response of these detectors to cosmic proton (ion) spectrum is given by the local geomagnetic cutoff 0.4 GV. This low cutoff is due to the Brazilian magnetic anomaly which has a magnetic field
minimum at 26S, 53W, which is very close to the position where the Tupi experiment is located. The magnetic minimum at ground in this region of the anomaly is three times lower than the magnetic field at the polar regions.  Thus, this area is like a funnel for incoming charged particles from space. 

The directionality of the vertical muon telescope is guaranteed by a veto or anti-coincidence guard, using a detector of the inclined telescope and vice-verse. Therefore, only muons with trajectories close to the telescope axis are registered. 
Every telescope has an effective field of view of 0.38 sr.
The data acquisition is made on the basis of the Advantech PCI-1711/73 card
with an analogical to digital conversion at a rate of up to 100 kHz. The Tupi experiment has a fully independent power supply, with an autonomy of up to 6 hours to guard against eventual power failures. As a result, the data acquisition is carried out during 24 hours, giving a duty cycle higher than 99\%.

\section{Learning with the EGRET GRB detections near bright BATSE bursts}

EGRET on the Compton Gamma Ray Observatory had observed several high energy and hour-long GRB triggers \cite{hurley94}.
These bursts had the highest fluence as recorded by BATSE within EGRET's field of view. The strongest detection was the GRB of 17 February 1994 with emissions at nearly 20 GeV and with durations over ten times longer than that detected by BATSE at MeV range. At least there were four long EGRET GRBs \cite{schneid92,sommer93,hurley95,schneid95} detected simultaneous with the bright BATSE sub-MeV emissions, and one of them is shown in Fig.2.

\begin{figure}[th]
\vspace*{-1.5cm}
\includegraphics[clip,width=0.5
\textwidth,height=0.5\textheight,angle=0.] {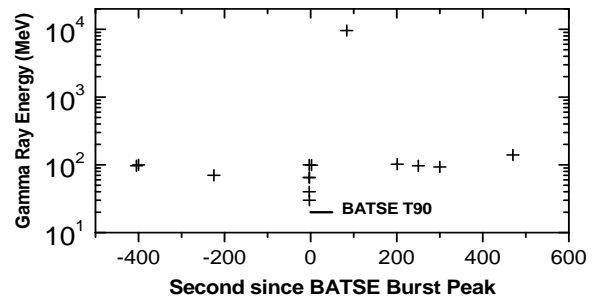}
\vspace*{-6.5cm}
\caption{The highest energy EGRET gamma ray detection near bright sub-MeV BATSE burst.}
\end{figure}

The above signature of long duration EGRET GRBs, with large time differences with BATSE
could be explained by shock front hitting an interstellar medium leading to sub-TeV to TeV gamma rays interacting with the infrared and cosmic microwave background radiations in the interstellar space, producing different path lengths from the source to the observer and a dispersion on the arrival time to the detector. 

The GRAND team, through a nice analysis has explored the possibility of observing GRBs with space borne detectors conjugated simultaneously with muon excess on the ground \cite{pourier03}. This implies that GeV to TeV counterpart photons would be an extension of the sub-MeV energies of the GRB spectrum. Whether these events could be explained by a simple extension of the low energy burst spectrum is not guaranteed due to the degradation of high energy burst during their propagation. There are EGRET GRBs with $\sim 100$ MeV to GeV gamma rays outside the interval T90 during which BATSE detects 90\% of the low energy flux.

Monte Carlo calculations have shown that a 1 TeV photon incident at the top of the atmosphere yields on the average 0.2 muons. Therefore, a muon enhancement could be expected at ground level only from a long GeV to TeV burst. According to the EGRET results, high energy ($\sim 100 MeV$) and long GRBs occur about one per year \cite{dingus97}. However, this rate must be higher, because the EGRET spark chamber detector had only an effective area of $1500 cm^2$ giving an effective angular openning of 18 degrees.

In this work, the search for GeV-TeV counterparts in GRBs, through their possible ground level muon enhancement is made on the basis of the long MeV EGRET GRBs signature in association with their sub MeV BATSE counterpart (see Fig.2), as following: 
firstly, we look for Tupi muon excess counting rate with a significance level above $4\sigma$, secondly, it is verified if the field of view of the telescope where there is the muon excess, include  a GRB coordinates, SWIFT and/or INTEGRAL, according to the GCN report, in coincidence or in 0.5 hours around the GRB occurrence.

\section{Results}

Starting from April of 2007 we have initiated the Phase II of the Tupi experiment with a survey on the daily variations of the muon intensity at sea level using two identical muon telescopes as described in section 2. The method applied here to study the cosmic ray anisotropy is based on the idea that a fixed detector scans the sky due to the Earth's rotation. Preliminary results of this survey can be found elsewhere \cite{navia07}.
Under the assumption of an isotropic distribution of GRB in the sky, the probability of finding the coordinates of a GRB inside the view of a telescope is 3\%. As now we are working with two telescopes, this proability is 6\%. 
Since April 2007, four SWIFT GRBs with their coordinates inside the field of view of the telescopes have been found. In particular, two of them have good confidence level. One has a confidence level of $4.4\sigma$ and is presented here. The other is of $3.2\sigma$.

\begin{figure}[th]
\vspace*{-1.15cm}
\includegraphics[clip,width=0.5
\textwidth,height=0.5\textheight,angle=0.] {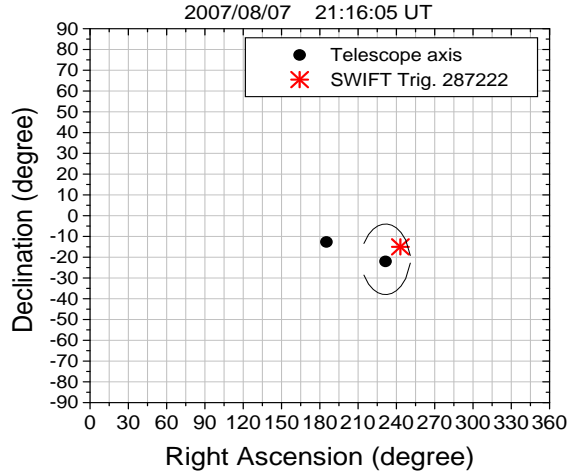}
\vspace*{-4.0cm}
\caption{Equatorial coordinates of the two Tupi telescope axes at SWIFT trigger time, with the ``circle'' representing the effective field of view (0.38 sr) of the Tupi vertical telescope, as well as the GRB position.}
\end{figure}

According to the CGN report \cite{barthelmy}, on 2007 August 07 at 21:16:05 UT the Burst Alert Telescope (BAT) on board of the SWIFT spacecraft had reported the arrival of an unknown event, Trigger 287222. Initially, the event was classified by SWIFT as a bright burst with a trigger duration of 
$64$ seconds in the energy band of 15-50 keV. The Swift-BAT event Position is RA:243.104d (+16h 12m 25s), and DEC:-15.019d. The rate significance and the background (intensity, time and duration) are undefined and no light curve was reported. In addition, the Swift StarTracker had lost stellar lock minutes bofore that
and the resulting improper s/c attitude information caused BAT
to "trigger" on a known source (in the apparent wrong position), so it is possibly bogus..
What of special has this unknown event?, the answer is that the coordinates of this event are
inside of the effective field of view of the Tupi vertical telescope as it is shown in Fig.3, where the two telescope axis equatorial coordinates are indicated together with the GRB coordinates. In addition, the effective field of view of the vertical telescope is added as a almost circle.

Figure 4 shows the telescope output (raw data) representing the 10 seconds muon counting rate. It is possible to see that there is a
GLE with a peak at 21:15:59 in coincidence with the SWIFT trigger occurrence within the 10-second counting interval. The GLE peak has a statistically significance of 4.4$\sigma$ and a duration of $\sim 10$ second. The GLE peak in Fig.4 is close to transition between an high and a low muon intensity
background \cite{navia07}. The effect is attributed to the corotating galactic cosmic ray with the interplanetary magnetic field (IMF) lines producing muons in the Earth atmosphere and it is responsible for the so called daily variation, the effect varies with the solar cycle.

\begin{figure}[th]
\vspace*{-1.0cm}
\includegraphics[clip,width=0.5
\textwidth,height=0.5\textheight,angle=0.] {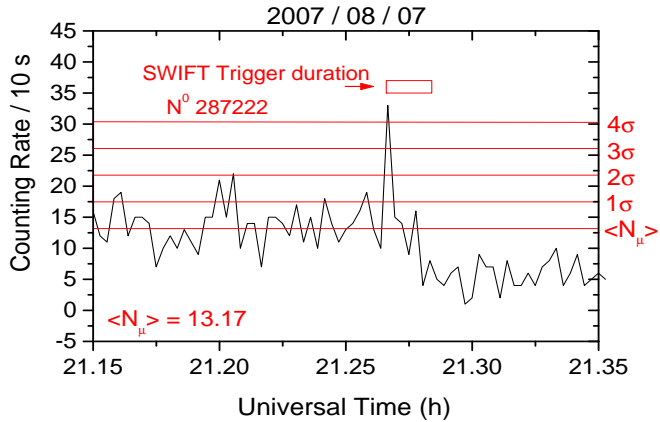}
\vspace*{-6.0cm}
\caption{The Tupi muon 10-second counting rate on 07 August 2007.}
\end{figure}

\section{Conclusions ans Remarks}

Despite the fact that gamma rays in the energy band of 30 GeV to several TeV have only 0.5\%-3.0\% chance of undergoing interactions in the atmosphere yielding pions, they are more efficient at producing energetic forward-directed pions. The $\gamma-Air$ interaction has a larger fraction of high energy pions than the $p-Air$ interaction \cite{fasso01}. In addition, Monte Carlo calculations (FLUKA) have shown that the average height above sea level at which the detected muons (in the GeV energy region) are produced is around 20 km for proton showers and 12 km for $\gamma$ showers. This means that the chance of ``photomuons'' reaching the ground is higher than the muons produced in $proton-Air$ interaction. These characteristics encourage the use of ground based method to detect GRBs. In order to give credibility to this technique, the detection of a GRB at ground level in coincidence with satellite detectors is necessary, at least in a firts step. 

We have presented here a GLE with strong signature of being associated to a SWIFT unknown event, Trigger 0287222, according to the criteria defined in section 3. They were obtained looking the signatures of 100 MeV EGRET GRBs in association with sub MeV BATSE counterpart. Unhappily, this trigger occured while the StarTracker had lost lock, so it is possibly bogus.  It may still be possible that this unknown event was produced by a Galactic source. 
 
We are waiting the next round of GRB satellites such as the GLAST, it will operate in the GeV region  and probably the chances of detection of GRBs at
ground and in coincidence with the satellite increase. In addition, there is the probability that in one year, the number of Tupi telescopes increases to eight or ten. 

So far, the GRBs study has shown an obvious advantage in the study of cosmology \cite{ghisellini05}, because they can be detected with redshift up to 10,
contrary to Ia Supernovae, detectable up to $z\sim 1.7$, and on the microwave background radiation, because it has an associated redshift (not measured) of $z\sim 1000$.  The SWIFT is detecting among 100 to 150 GRBs per year and in most of cases with redshift
and light curves very well established. Allowing the most optimistic expectation for doing cosmology with GRBs.

\section{Acknowledments}

This work is supported by the National Council for Research (CNPq) of Brazil, under Grant No. $479813/2004-3$. The authors wish to express their thanks to Dr. A. Ohsawa of Tokyo University for help in the first stage of the experiment. We want to thank to Alexander Kann, Kim Page, Nicola Masetti and Evert Rol to alert us of the conditions of the SWIFT trigger 0287222 and to S. Barthelmy for additional informations.
We are also grateful to various catalogs available on the web and to their open data policy, especially to the GCN report.


\newpage

\end{document}